\documentclass[journal=jacsat,manuscript=article]{achemso}

\usepackage{chemformula} 
\usepackage[T1]{fontenc} 
\usepackage{graphicx}
\usepackage{soul} 
\usepackage{bm}
\usepackage{amsfonts}
\graphicspath{
{./figure/}
{./}
}
\usepackage{lineno}

\author{Kairi Masuda}
\email{kairi.masuda.c5@tohoku.ac.jp}
\author{Yu Kumagai}

\affiliation[Tohoku University]
{Institute for Materials Research, Tohoku University, 2-1-1 Katahira, Aoba-ku, Sendai, 980-8577, Japan}

\title[An \textsf{achemso} demo]
{Atomic-scale phase-field modeling \\ for 2D ferroelectrics\\including non-Gaussian fluctuations}

\abbreviations{IR,NMR,UV}

\begin{document}






\begin{abstract}
Atomic-scale phase-field modeling extends the phase-field framework down to the level of individual atoms by treating the probability density of atomic vibrations as a field variable and constructing a corresponding free-energy functional from this atomic field together with interatomic potentials. In this way, the framework has the potential to visualize local thermodynamic states with atomic-level resolution, just as conventional phase-field modeling has served as a computational microscope for free energy, stress, and related quantities at mesoscopic scales. However, existing formulations mainly assume Gaussian probability distributions for atomic vibrations, which limits their applicability to more complex and heterogeneous systems such as surfaces. In this work, we generalize the atomic-scale phase-field methodology by extending the free-energy functional to include non-Gaussian fluctuations. We apply this approach to monolayer SnTe in the NVT ensemble and show that the predicted equilibrium polarization is in better agreement with molecular dynamics simulations and that the ferroelectric-to-paraelectric phase transition is successfully reproduced. Furthermore, by decomposing the entropy on a per-atom basis, we visualize atomically resolved maps of local entropy and find that Sn atoms contribute more strongly to the entropy than Te atoms at high temperature, which is a driving factor of the ferroelectric-to-paraelectric phase transition. These results broaden the applicability of phase-field approaches to a wider range of atomic systems and suggest a route toward an atom-resolved theory of phase transitions based on high-resolution thermodynamics.

\end{abstract}


\clearpage

\section{Introduction}
In phase-field modeling\mbox{\cite{LQchenReview,SolidificationReview,Ambati2015,doi:10.1098/rsta.2015.0166}}, a target phenomenon is represented by a field variable, and the free energy, which is typically treated as a global property in other methodologies such as molecular dynamics\mbox{\cite{FrenkelLadd,BENNETT1976245,BONOMI20091961}}, is formulated as a position-dependent functional of that field. The system then evolves to minimize this spatially inhomogeneous free-energy landscape. This local thermodynamic framework has proved highly effective for simulating ferroelectric materials based on Landau-Ginzburg-Devonshire (LGD) theory\mbox{\cite{doi:10.1080/14786444908561372,doi:10.1080/14786445108561354}}, which has successfully reproduced various domain states and topological phases, such as polarization vortices and skyrmions\mbox{\cite{LQchen_review_thin_film,doi:10.1038/nature16463,doi:10.1038/s41586-019-1092-8,doi:10.1038/s41563-020-0694-8,doi:10.1038/s41467-023-36950-x,doi:10.1126/science.1259869}}. Furthermore, phase-field simulations not only reproduce such inhomogeneous polarization distributions but also provide spatially resolved thermodynamic information, including local free energy, stress, and entropy\mbox{\cite{CHOUDHURY20055313,Gao2013,Park2018,WANG2021117383,NatureYang2025,doi:10.1126/science.abb3209}}.
As a result, phase-field modeling has served as a computational microscope for visualizing local thermodynamic states in materials systems. On the other hand, recent breakthroughs in materials science have shifted attention toward smaller length scales, where heterogeneity extends down to individual atoms. Examples include the discovery of various grain-boundary phases\mbox{\cite{doi:10.1126/science.adq4147,doi:10.1038/s41586-020-2082-6,PhysRevLett.120.267601}}, atomic-scale real-time observation of polarization switching\mbox{\cite{doi:10.1126/science.adh7670}}, the emergence of high-entropy alloys/ceramics\mbox{\cite{MIRACLE2017448,doi:0.1038/ncomms9485,doi:10.1126/science.adl2931}}, and renewed interest in molecular ferroelectrics\mbox{\cite{Horiuchi2008,elastin,doi:0.1038/s41467-021-21019-4,doi:10.1038/36069}}. However, phase-field modeling, being rooted in continuum mechanics, cannot be directly applied to such atomistic systems.

To address this need, the phase-field crystal concept has been proposed, in which a crystallite is treated as a field object and the resulting free-energy functional successfully reproduces crystal growth and dislocation motion\mbox{\cite{doi:10.1103/PhysRevE.70.051605,doi:10.1103/PhysRevE.79.035701,doi:10.1103/PhysRevE.73.031609,doi:10.1007/s11837-007-0095-3,doi:10.1103/PhysRevMaterials.4.013802}}. To further increase the thermodynamic resolution, the probability density of atomic vibrations is treated as a single field object, and a free-energy functional can be constructed based on this atomic field and interatomic potentials\mbox{\cite{PhysRevLett.63.624,doi:10.1063/1.460547,PhysRevB.84.054103,PhysRevB.110.104107,doi:10.1103/ngky-crt4}}; in this study, we refer to this approach as atomic-scale phase-field modeling. This formulation may enable the free energy to be decomposed into contributions from individual atoms, thereby providing single-atom-level insight into local thermodynamic states. This theory is further extended to include many-body interactions with the aid of the self-consistent harmonic approximation (SSCHA)\mbox{\cite{PhysRevB.89.064302,PhysRevB.98.024106,doi:10.1038/s41586-020-1955-z,doi:10.1088/1361-648X/ac066b,PhysRevB.110.144101}} and, consequently, machine learning potentials\mbox{\cite{10.1063/1.4812323,SCHMIDT2024101560,doi:10.1021/acscatal.2c05426,Zeni2025,Batatia2025,Deng2023,Takamoto2022}}.

Despite these advances, most existing implementations rely on a Gaussian description of atomic fluctuations, in which atomic vibrations are assumed to be harmonic for the sake of theoretical simplicity. This approximation is often reasonable for bulk crystals, but it can become inadequate in systems with surfaces or reduced dimensionality, where broken local symmetry and reduced coordination can lead to skewed, non-Gaussian atomic distributions. This issue is especially important for 2D ferroelectrics, which have attracted significant attention due to their technological and scientific importance\mbox{\cite{doi:10.1126/science.aad8609,doi:10.1126/science.abd3230,doi:10.1126/science.abe8177,doi:10.1038/s41586-020-2970-9}}. Therefore, an atomic-scale phase-field framework that incorporates non-Gaussian fluctuations is needed to extend this approach to low-dimensional ferroelectric systems. Such a framework would enable thermodynamic quantities to be visualized with atomic resolution across ferroelectric phase transitions, thereby providing direct insight into the atomic-scale mechanisms of phase stability.

Here, we develop atomic-scale phase-field modeling incorporating non-Gaussian fluctuations for broader applicability. This paper is organized as follows. First, we extend the free-energy functional of atomic systems to include non-Gaussian fluctuations and describe the computational details of the calculations. Next, we apply the proposed methodology to a 2D ferroelectric material, namely monolayer SnTe, and validate the theory by comparing the predicted electric polarization with molecular dynamics simulations. We then describe how to decompose the entropy on a per-atom basis for visualization and clarify the underlying mechanism of the ferroelectric-paraelectric phase transition in SnTe with atomic resolution. Finally, we summarize the key findings and discuss the future prospects of this approach.

\clearpage
\section{Methods}

\subsection{Theory}
We consider atomic vibrations governed by the following anharmonic potential:
\begin{eqnarray}
U(\bm{x})&=&\frac{1}{2}k_{ij}x_{i}x_{j}+\frac{1}{3}k_{ijk}x_{i}x_{j}x_{k}+\frac{1}{4}k_{ijkl}x_{i}x_{j}x_{k}x_{l},
\end{eqnarray}
where $k_{ij}$, $k_{ijk}$, and $k_{ijkl}$ are force constants, and $x_{i}$ denotes the displacement. Furthermore, to ensure positive definiteness of the harmonic term, the matrix $\boldsymbol{k}=(k_{ij})$ is parameterized through its eigendecomposition as follows:
\begin{eqnarray}
\boldsymbol{k}=\bm{R} \bm{\Lambda} \bm{R}^{T}=\bm{R}\begin{pmatrix}
\lambda_{x'} & 0  & 0 \\
0 & \lambda_{y'}  &  0  \\
0 & 0 & \lambda_{z'}
\end{pmatrix}\bm{R}^{T},
\end{eqnarray}
where the eigenvalues $\lambda_{x'}$, $\lambda_{y'}$, and $\lambda_{z'}$ are constrained to be positive, and $x'$, $y'$, and $z'$ denote the principal axes of $\bm{k}$. 
Here, $\bm{\Lambda}=\mathrm{diag}(\lambda_{x'}, \lambda_{y'}, \lambda_{z'})$ is the diagonal matrix of eigenvalues, and $\bm{R}$ is a three-dimensional rotation matrix, which can be decomposed into rotation matrices about each axis as follows:
\begin{eqnarray}
\bm{R}= \bm{R}_{z}(\theta_{z})\bm{R}_{y}(\theta_{y})\bm{R}_{x}(\theta_{x}).
\end{eqnarray}
Therefore, an anharmonic potential is specified by 31 variables, that is, $\bm{\Sigma}$ = ($\lambda_{x'}$, $\lambda_{y'}$, $\lambda_{z'}$, $\theta_{x}$, $\theta_{y}$, $\theta_{z}$, $k_{ijk}$, and $k_{ijkl}$), and the corresponding probability distribution of atomic vibrations is given by
\begin{eqnarray}
\rho(\boldsymbol{r}|\boldsymbol{X},\bm{\Sigma})&=&\frac{e^{-\beta U(\bm{r}-\bm{X})}}{Z}.
\end{eqnarray}
Here, $\bm{x}=\bm{r}-\bm{X}$, where $\bm{r}$ is the position of the oscillator and $\bm{X}$ is its mean position. Also, $\beta=1/(k_{B}T)$, where $k_{B}$ and $T$ are the Boltzmann constant and temperature, respectively. $Z$ is the potential-energy part of the partition function, given by
\begin{eqnarray}
Z=\int e^{-\beta U(\bm{r}-\bm{X})} d\boldsymbol{r}.
\end{eqnarray}
Using the partition function $Z_{\rm{ah}}= Z_{\rm{kin}}Z$, where $Z_{\rm{kin}} = \Big(\frac{m k_{B}T}{2\pi \hbar^2}\Big)^{\frac{3}{2}}$ is the kinetic part, the corresponding free energy of anharmonic oscillation is calculated as follows (the derivation is given in Supporting Information S1):
\begin{eqnarray}
F_{\rm{ah}}&=&-k_{B}T\ln Z_{\rm{ah}} \\ \nonumber
&=&-k_{B}T \ln \Big(\frac{m k_{B}T}{2\pi \hbar^2}\Big)^{\frac{3}{2}} -k_{B}T \ln \Bigg\{\det|\bm{\Lambda}^{-\frac{1}{2}}|\Big(\frac{2\pi}{\beta}\Big)^{\frac{3}{2}} \mathbb{E}_{\bm{z}}[w] \Bigg\}, \\ \nonumber
\end{eqnarray}
where $\hbar$ is the Planck constant, $m$ is the atomic mass, and $\det|\bm{\Lambda}^{-\frac{1}{2}}|$ denotes the determinant of $\bm{\Lambda}^{-\frac{1}{2}}$. $\mathbb{E}_{\bm{z}}[w]$ is the expectation value with respect to the standard three-dimensional normal distribution, i.e., $\bm{z}\sim N(0,\bm{I})$, where $w$ is the weight function represented as
\begin{eqnarray}
w&=&\exp \Big(-\frac{1}{3}\beta k_{ijk} x_{i}x_{j}x_{k}-\frac{1}{4}\beta k_{ijkl} x_{i}x_{j}x_{k}x_{l} \Big) \\ \nonumber
&=&\exp \Big(-\frac{1}{3}\beta^{-\frac{1}{2}}k_{ijk}A_{ia}A_{jb}A_{kc}z_{a}z_{b}z_{c}-\frac{1}{4}\beta^{-1}k_{ijkl}A_{ia}A_{jb}A_{kc}A_{ld}z_{a}z_{b}z_{c}z_{d} \Big),
\end{eqnarray}
where we used the relation $\bm{x}=\beta^{-\frac{1}{2}}\bm{A}\bm{z}$ with $\bm{A}=\bm{R}\bm{\Lambda}^{-\frac{1}{2}}$.

The Gibbs-Bogoliubov inequality sets an upper bound on the free energy, i.e., $F$ as follows:
\begin{eqnarray}
F_{\rm{true}}\leq F_{0}+\langle \phi-V_{0}\rangle_{0} \equiv F,
\label{eq:GibbsBogo}
\end{eqnarray}
where $F_{\rm{true}}$ is the true free energy associated with the interatomic potential $\phi$, while $F_{0}$ is a reference free energy associated with the reference potential energy $V_{0}$. $\langle \cdot \rangle_{0}$ denotes an expectation value with respect to the reference probability distribution governed by $V_{0}$. LeSar \textit{et al.} employed an Einstein solid as the reference state and derived an analytical form of $F$ using mutually independent isotropic Gaussian\mbox{\cite{PhysRevLett.63.624,doi:10.1063/1.460547,PhysRevB.84.054103,PhysRevB.110.104107}} for Cu described by a Morse pair potential. Here, we extend their theory to non-Gaussian distributions by using the above anharmonic potential of $N$ atoms as the reference state, i.e., $F_{0}=\sum_{n} F_{ah,n}$ and $V_{0}=\sum_{n} U_{n}$, together with many-body interactions, as follows:
\begin{eqnarray}
\label{eq:freeene}
F&=&-k_{B}T\sum_{n=1}^{N} \ln \Big(\frac{m_{n} k_{B}T}{2\pi \hbar^2}\Big)^{\frac{3}{2}} -k_{B}T\sum_{n=1}^{N} \ln \Bigg\{\det|\bm{\Lambda}_{n}^{-\frac{1}{2}}|\Big(\frac{2\pi}{\beta}\Big)^{\frac{3}{2}} \mathbb{E}_{\bm{z}_{n}}[w_{n}] \Bigg\} \\ \nonumber
&&+\int\int...\int{\rho_{1}(\boldsymbol{r}_{1})}{\rho_{2}(\boldsymbol{r}_{2})}...{\rho_{N}(\boldsymbol{r}_{N})}\phi(\boldsymbol{r}_{1},\boldsymbol{r}_{2},...,\boldsymbol{r}_{N})d\boldsymbol{r}_{1}d\boldsymbol{r}_{2}...d\boldsymbol{r}_{N} \\ \nonumber
&&-\int\int...\int{\rho_{1}(\boldsymbol{r}_{1})}{\rho_{2}(\boldsymbol{r}_{2})}...{\rho_{N}(\boldsymbol{r}_{N})} \Bigg(\sum_{n=1}^{N} U_{n} (\boldsymbol{r}_{n}-\boldsymbol{X}_{n}) \Bigg) d\boldsymbol{r}_{1}d\boldsymbol{r}_{2}...d\boldsymbol{r}_{N}. \nonumber
\end{eqnarray}
The third and fourth terms are the expectation values of the many-body potential energy $\phi$, i.e., an effective interaction among the probability clouds, and of the reference potential energy, respectively, corresponding to $\langle \phi- V_{0} \rangle_{0}$. The true free energy $F_{\rm{true}}$ is approximated variationally by minimizing $F$ with respect to the shape parameters $\bm{\Sigma}_{n}$  of the anharmonic probability densities.


Under given positions $\boldsymbol{X}_{n}$ and the optimal parameters of the anharmonic potential $\boldsymbol{\Sigma}_{n}$ calculated above, the positions should evolve in time so as to reduce the free energy while conserving the atomic probability density, i.e., $\int \rho \, d\boldsymbol{r} = 1$. This implies that the probability density should evolve according to a conservative phase-field equation as follows:
\begin{eqnarray}
\label{eq:govern}
  \frac{\partial \rho_{n}(\textbf{\textit{r}},t)}{\partial t}&=&\nabla \cdot \left(D_{n}(\textbf{\textit{r}})\nabla \frac{\delta F}{\delta \rho_{n}} \right) \\ \nonumber
  &=& \nabla \cdot \left(\kappa_{n} \rho_{n}(\textbf{\textit{r}}) \nabla \Phi_{n}(\textbf{\textit{r}})\right),
\end{eqnarray}
where $t$ is time and $D_{n}(\textbf{\textit{r}})$ is a position-dependent kinetic coefficient. Here, since $\kappa_{n}$ is a kinetic coefficient, we set $D_{n}(\textbf{\textit{r}}) = \kappa_{n} \rho_{n}(\textbf{\textit{r}})$ so that the probability density satisfies the continuity equation. Here, the gradient of the functional derivative, $\nabla (\delta F/\delta \rho_{n}) = \nabla \Phi_{n}(\textbf{\textit{r}})$, acts as an effective force. By solving the above governing equation, the motion of the probability densities can be calculated.

\clearpage

\subsection{Simulation detail}
To validate the proposed theory, we selected monolayer SnTe, which has been shown to be a 2D ferroelectric, as a model system, and evaluated its pressure evolution in the canonical (NVT) ensemble. A conventional-cell-based 2 $\times$ 2 $\times$  1 supercell with lattice constants of $a$ = 4.64 \mbox{\AA}, $b$ = 4.56 \mbox{\AA}, and $c$ = 3.16 \mbox{\AA} was constructed, and a vacuum region of 10 \mbox{\AA} was attached on both sides along the $z$ direction, resulting in a 16-atom supercell with dimensions of 9.28 \mbox{\AA} $\times$ 9.12 \mbox{\AA} $\times$ 23.16 \mbox{\AA}\mbox{\cite{doi:10.1038/s41524-024-01213-w}}.



Figure 1 presents a schematic overview of the computational workflow. The third term in the free-energy functional in Eq.~(\mbox{\ref{eq:freeene}}) is a high-dimensional integral that is most efficiently evaluated by Monte Carlo techniques. Because the transformed variables $\bm{z}_{n}$ follow mutually independent standard normal distributions, we employ importance sampling: random configurations are drawn from these Gaussian distributions rather than from a uniform distribution. That is, we evaluate the integral as follows (a derivation is given in Supplemental Information S1):
\begin{eqnarray}
\label{eq:MonteCarloIntegral}
\langle \phi \rangle_{0}&=&\int \cdots \int \rho_{1}(\boldsymbol{r}_{1}) \cdots \rho_{N}(\boldsymbol{r}_{N}) \phi(\boldsymbol{r}_{1},\ldots,\boldsymbol{r}_{N})\, d\boldsymbol{r}_{1}\cdots d\boldsymbol{r}_{N} \\ \nonumber
&=&\mathbb{E}_{\bm{z}_1,\ldots,\bm{z}_N} \Big[\phi(\bm{X}_1+\beta^{-1/2}\bm{A}_{1} \bm{z}_1,\ldots,\bm{X}_N+\beta^{-1/2}\bm{A}_{N} \bm{z}_N)\, W'(\bm{z}_1,\ldots,\bm{z}_N)\Big],
\end{eqnarray}
where $W' = W/\mathbb{E}_{\bm{z}_1,\ldots,\bm{z}_N} [W]$ is the normalized total weight function, with $W = \prod_{n=1}^{N} w_{n}$. To reduce the statistical error of the Monte Carlo estimate, we employ a Sobol low-discrepancy sequence\mbox{\cite{SOBOL196786}} and map it to the Gaussian ensemble through the inverse cumulative distribution function, resulting in a total sample size of $N_{\rm{sample}}=1000$. The interatomic potential $\phi$ is provided by the Multi-Atomic Cluster Expansion (MACE) model\mbox{\cite{Batatia2025}}, which encodes each atomic configuration as a graph and is trained on r$^2$SCAN data\mbox{\cite{doi:10.1021/acs.jpclett.0c02405,kaplan2025foundationalpotentialenergysurface}}. Because MACE can evaluate many graphs concurrently, the integrand is computed in parallel on a GPU via PyTorch\mbox{\cite{paszke2017automatic}}, allowing the high-dimensional integral to be evaluated with minimal wall-time overhead. In the same manner, we evaluate $\langle V_{0} \rangle_{0}$ for the fourth term in the free-energy functional in Eq.~(\mbox{\ref{eq:freeene}}).

The gradients of the free energy with respect to the probability-density shape parameters $\bm{\Sigma}_{n}$ for atom $n$ are evaluated using the following equations (see Supplementary Information S2 for the detailed expressions):
\begin{eqnarray}
\frac{\partial F}{\partial \lambda_{n,i}}&=&\frac{\partial F_{0}}{\partial \lambda_{n,i}}+\frac{\partial \langle \phi \rangle_{0}}{\partial \lambda_{n,i}}-\frac{\partial \langle V_{0} \rangle_{0}}{\partial \lambda_{n,i}},
\end{eqnarray}
\begin{eqnarray}
\frac{\partial F}{\partial \theta_{n,i}}&=&\frac{\partial F_{0}}{\partial \theta_{n,i}}+\frac{\partial \langle \phi \rangle_{0}}{\partial \theta_{n,i}}-\frac{\partial \langle V_{0} \rangle_{0}}{\partial \theta_{n,i}},
\end{eqnarray}
\begin{eqnarray}
\frac{\partial F}{\partial k_{n,ijk}}&=&\frac{\partial F_{0}}{\partial k_{n,ijk}}+\frac{\partial \langle \phi \rangle_{0}}{\partial k_{n,ijk}}-\frac{\partial \langle V_{0} \rangle_{0}}{\partial k_{n,ijk}},
\end{eqnarray}
\begin{eqnarray}
\frac{\partial F}{\partial k_{n,ijkl}}&=&\frac{\partial F_{0}}{\partial k_{n,ijkl}}+\frac{\partial \langle \phi \rangle_{0}}{\partial k_{n,ijkl}}-\frac{\partial \langle V_{0} \rangle_{0}}{\partial k_{n,ijkl}}.
\end{eqnarray}
The subsequent minimization of the free-energy functional is carried out using the 
Limited-memory Broyden–Fletcher–Goldfarb–Shanno (L-BFGS) algorithm implemented in SciPy\mbox{\cite{LBFGS,2020SciPy-NMeth}}, with a relative convergence tolerance of $1.0 \times 10^{-6}$.


Our many-body free-energy framework and the corresponding numerical procedure share conceptual similarities with the stochastic self-consistent harmonic approximation (SSCHA), which also employs importance sampling within a variational formulation to evaluate anharmonic phonon free energies\mbox{\cite{PhysRevB.89.064302,PhysRevB.98.024106,doi:10.1038/s41586-020-1955-z,doi:10.1088/1361-648X/ac066b,PhysRevB.110.144101}}. A key difference lies in the choice of variational parameters. In the present work, we optimize the probability density directly in real space, introducing $O(N)$ independent variables for a system of $N$ atoms, whereas SSCHA determines the full auxiliary dynamical matrix, which in general involves $O(N^2)$ parameters. The substantially reduced parameter space makes our method more amenable the explicit treatment of defects and other local perturbations.



For the phase-field equation, Eq. (\mbox{\ref{eq:govern}}) has the same form as the Fokker-Planck equation described as follows:
\begin{equation}
  \frac{\partial \rho_{i}(\boldsymbol{x},t)}{\partial t}=\nabla ( \boldsymbol{a}_{i}(\boldsymbol{x}) \rho_{i}(\boldsymbol{x})),
\end{equation}
where $\boldsymbol{a}_{i}$ is a drift coefficient. To simplify this equation, we assume that each probability-density distribution translates without deformation during an infinitesimal motion, and that the distribution shape is then updated. Under this assumption, the equation can be reduced to the following Langevin equation\mbox{\cite{PhysRev.36.823,PhysRevB.110.104107}}:
\begin{equation}
\label{eq:langevin}
  \frac{d\boldsymbol{X_{i}}}{dt}=-\kappa_{i} \nabla_{i} F(\boldsymbol{X_{i}}).
\end{equation}
We calculate the motion of the probability densities by solving this equation using the Verlet algorithm with a time step of $\Delta t = 100$ fs. The derivative of the free energy with respect to $X_{n,i}$ can be written as follows (a derivation is given in Supplemental Material S3):
\begin{eqnarray}
 -\nabla F &=&-\frac{d F}{d X_{n,i}} \\ \nonumber
 &=&-\mathbb{E}_{\bm{z}_{1},...,\bm{z}_{N}} \bigg[\frac{\partial \phi}{\partial X_{n,i}} W' \bigg]
\end{eqnarray}
Thus, the negative free-energy gradient, i.e., the mean force, is the expectation value of the force due to the interatomic potential. For the mobility $\kappa$, we employed values of $2.98 \times 10^{-8}$ for Sn and $2.77 \times 10^{-8}$ for Te, with units of m$^{2}$/(V$\cdot$ s). These values were estimated from $\kappa=5.56\times10^{-8}$ m$^{2}$/(V$\cdot$ s), which is the mobility of Cu$^{2+}$ in water at 298 K\mbox{\cite{Cumobility}}, and the relation $\kappa = |q|/(m v_{m})$, where $q$ is the charge, $m$ is the atomic mass, and $v_{m}$ is the collision frequency\mbox{\cite{Ion_mobility_equation}}.

For comparison, we conducted the ordinal molecular dynamics (MD) simulation using the Atomic Simulation Environment (ASE) to calculate the time evolution of the above initial structure under NVT and NPT ensembles\mbox{\cite{ase-paper,ISI:000175131400009}} of 40000 steps with the time step being 5 fs.



 
 \begin{figure}
  \includegraphics{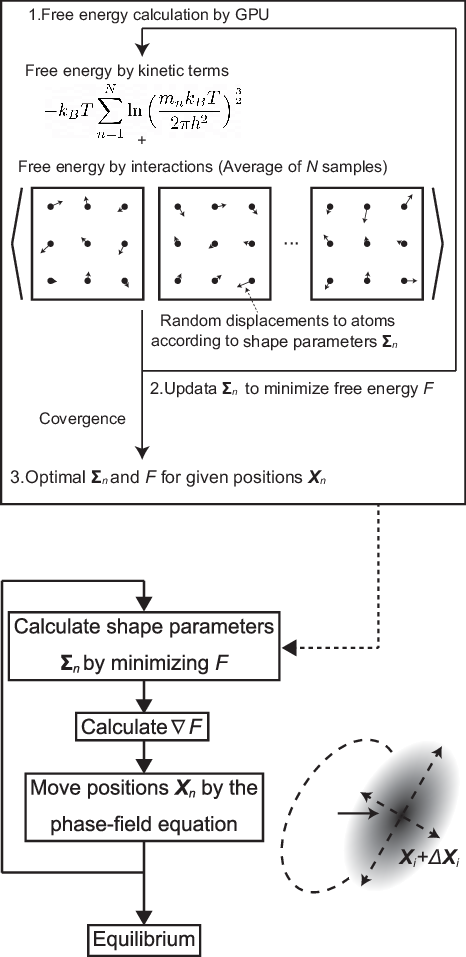}
  \caption{A schematic illustration of the computational workflow for phase-field modeling that includes many-body interactions.}
  \label{fig:nvttime}
\end{figure}

\clearpage

\section{Results}
\subsection{Time evolution}
\begin{figure}
  \includegraphics{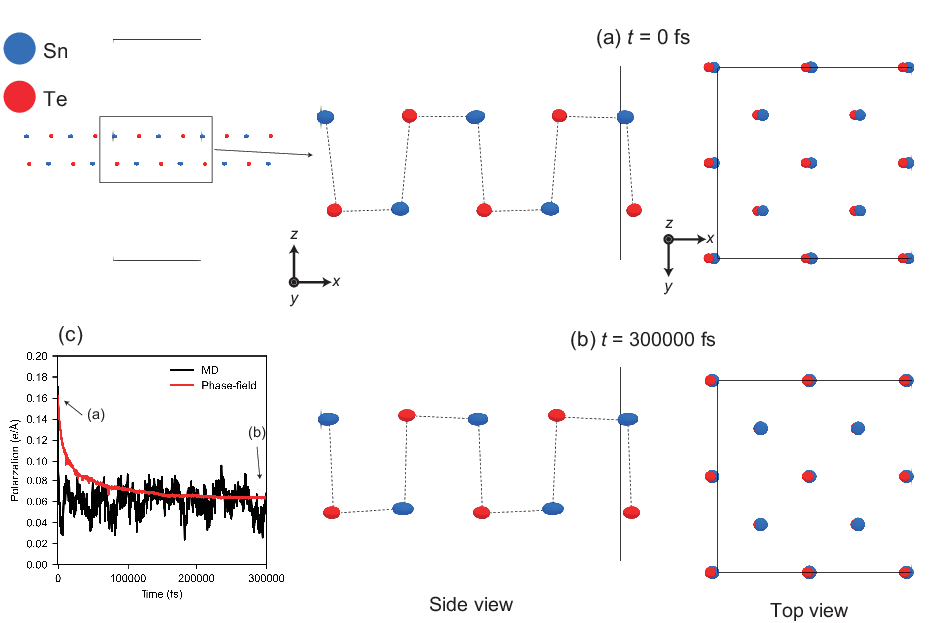}
  \caption{(a,b) Time evolution of the atomic probability density distribution, characterized by $\boldsymbol{\Sigma}_{i}$, for Sn and Te atoms obtained by phase-field simulations in the NVT ensemble at $T = 300$ K. The $x$, $y$, and $z$ axes correspond to the crystallographic $a$, $b$, and $c$ directions, respectively. Note that the dispersion of the atomic density is shown as an isosurface enclosing 68\% of the probability mass. For an isotropic Gaussian distribution, this corresponds to the 1$\sigma$ interval in one dimension. Furthermore, the dispersion is magnified by a factor of 1.5. (c) Time evolution of the polarization obtained by the phase-field and MD simulations.}
  \label{fig:time}
\end{figure}

Figures \mbox{\ref{fig:time}}(a) and (b) and the Supplemental Material show the time evolution of the atomic probability densities in monolayer SnTe in the NVT ensemble at temperature $T = 300$ K, obtained by phase-field simulations. As the initial structure, we consider a ferroelectric state obtained by structural minimization of the atomic positions without finite-temperature effects. Note that the polarization direction in the low-temperature phase is along the $x$-axis. We found that Sn and Te atoms, which are initially displaced from symmetric positions, move toward high-symmetry positions. This is consistent with the experimental observation that monolayer SnTe undergoes a phase transition at $T = 270$ K from the ferroelectric state to the paraelectric state\mbox{\cite{doi:10.1126/science.aad8609}}. We then calculated the time evolution of the polarization using the following equation (the derivation is given in Supplemental Material S5):
\begin{eqnarray}
\bm{P}^{2D}&=&-\frac{1}{S}\frac{d F}{d \bm{E}}\bigg|_{\bm{E}=0} \\ \nonumber
&=&\Big\langle \frac{e}{S} \sum_{j} \bm{Z}_{j} \bm{u}_{j} \Big\rangle_{0},
\end{eqnarray}
where $\bm{E}$ is an electric field, $S$ is the area of the system, $e$ denotes the elementary charge, and $\bm{u}_{j}$ is the displacement vector of atom $j$ from the paraelectric state. $\bm{Z}_{j}$ is the Born effective charge tensor, where $\bm{Z}_{\rm{Sn}} = \mathrm{diag}(6.23,6.23,6.23)$ and $\bm{Z}_{\rm{Te}} = \mathrm{diag}(-6.23,-6.23,-6.23)$\mbox{\cite{doi:10.1038/s41524-020-00449-6}}. Figure \mbox{\ref{fig:time}} shows the time evolution of the polarization calculated by the phase-field simulation (red line). For comparison, the polarization calculated by MD is also plotted (black line). We found that the polarization decreases with time and reaches an equilibrium value of $P_{x} = 6.37 \times 10^{-2}$ (e/\mbox{\AA}). This value is close to the mean polarization value of $P_{x} = 6.02 \times 10^{-2}$ (e/\mbox{\AA}) obtained by MD simulations. Therefore, our phase-field modeling successfully reproduces the polarization behavior at this temperature.

\clearpage


\clearpage

\clearpage
\subsection{Temperature dependence of polarization}

\begin{figure}
  \includegraphics{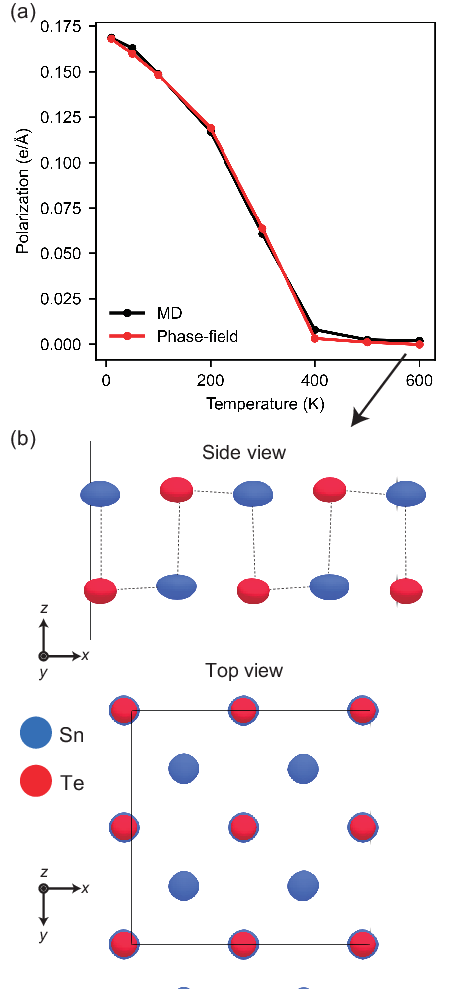}
  \caption{(a) Equilibrium polarization of monolayer SnTe obtained by phase-field and MD simulations in the NVT ensemble at various temperatures, and (b) the corresponding atomic probability density distribution at 600 K.}
  \label{fig:temperature}
\end{figure}

To investigate the temperature dependence of the polarization, we calculated the equilibrium polarization at various temperatures in the NVT ensemble. Figure \mbox{\ref{fig:temperature}}(a) shows the relationship between temperature and polarization obtained by phase-field and MD simulations. We found that the polarization values from both simulations agree with each other. Thus, our phase-field modeling can reproduce the temperature dependence of the polarization in this atomic system. On the other hand, when we focus on Sn atoms (blue atoms) at 600 K, we found that the atomic probability density extends toward the vacuum direction and deviates from a symmetric Gaussian distribution, indicating that the atomic vibration is no longer harmonic (Fig. \mbox{\ref{fig:temperature}}(b)). This suggests that our modeling successfully incorporates the anharmonicity of atomic vibrations at the surface.


On the other hand, it should be noted that polarization switching can occur because of thermal fluctuations during MD simulations near the transition temperature, i.e., around 300 K. To exclude this dynamical switching effect, we evaluated the polarization using the sample that did not show polarization switching during the simulation, as shown in Fig.~2(c). Therefore, the actual long-time-averaged polarization around 300 K is expected to be smaller than the value shown in Fig.~3(a). At this stage, our phase-field modeling cannot reproduce such switching dynamics because atomic fluctuations are modeled as a unimodal distribution. To incorporate such effects, it will be necessary to introduce a multimodal distribution, which remains for future work.

\clearpage

\clearpage
\subsection{Atom-resolved phase transition mechanism of SnTe}

A common atomistic explanation for phase transitions in ferroelectric materials is the soft-phonon scenario, in which a specific mode softens with decreasing temperature, becomes unstable, and freezes into a large-amplitude distortion\mbox{\cite{PhysRevLett.3.412,doi:10.1080/00018736000101229,doi:10.1080/00018736100101321}}. Since phonon modes are labeled in reciprocal space, however, extracting local real-space insight from them is not straightforward. On the other hand, the free energy constructed in the present study successfully reproduces the temperature dependence of the polarization. Based on this, we attempt to gain real-space insight into the phase-transition mechanism from the variation of the local free energy. To this end, we generated configurations spanning polarizations from 0 to 0.2 e/\mbox{\AA} by interpolating the atomic displacements obtained in the NVT simulations (Figs.~\mbox{\ref{fig:time}}(a) and (b)). We then conducted single-point calculations, i.e., we evaluated the free energy at various temperatures for these fixed configurations by optimizing only the probability densities of atomic vibrations under the stricter condition of $N_{\rm{sample}}=10000$.

\begin{figure}
  \includegraphics{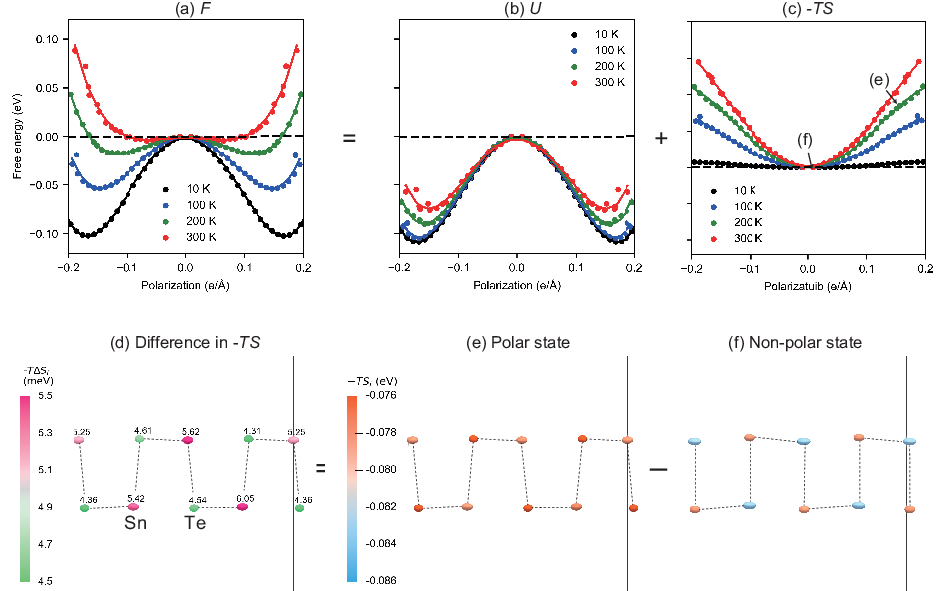}
\caption{(a) Free energy, (b) internal energy, and (c) entropic contribution ($-TS$) profiles as functions of polarization. (d) Difference in the entropic contribution, and (e) and (f) entropic-contribution maps for the polar and non-polar states at 200 K.}
  \label{fig:doublewell}
\end{figure}

Figure \mbox{\ref{fig:doublewell}}(a) plots the free energy as a function of polarization obtained from single-point calculations at $T = 10$--300 K. Here, the free energy is referenced such that it is zero at zero polarization. The resulting profiles exhibit a double-well potential, and the depth of the wells decreases with increasing temperature. This indicates that the temperature dependence of the polarization observed in our simulations originates from the gradual disappearance of the double-well minima, consistent with the standard LGD picture\mbox{\cite{doi:10.1080/14786444908561372,doi:10.1080/14786445108561354}}. Unlike the phenomenological LGD formulation, however, our free-energy landscape is obtained from interatomic potentials trained on first-principles calculations. Consequently, we can directly evaluate the entropy as follows (the derivation is given in Supporting Information S6):
\begin{eqnarray}
\label{eq:entropy}
S&=&-\bigg(\frac{\partial F}{\partial T}\bigg)_{V,N} \\ \nonumber
&=& \sum_{n=1}^{N} \bigg\{
 k_{B}\ln Z_{n} +k_{B}T \frac{\partial \ln Z_{n}}{\partial T} \\ \nonumber
&-&\mathbb{E}_{\{\bm{z}_{m}\}}\bigg[\nabla_{\bm{r}_{n}} \phi \cdot \frac{\bm{x}_{n}}{2 T} W' \bigg]-\mathrm{Cov}_{W'}\bigg[\phi, \frac{\partial \ln w_{n}}{\partial T} \bigg]  \\ \nonumber
&+&\mathbb{E}_{\{\bm{z}_{m}\}}\bigg[\frac{\partial U_{n}}{\partial T} W' \bigg]+\mathrm{Cov}_{W'}\bigg[U_{n}, \frac{\partial \ln W}{\partial T} \bigg] \bigg\},
\end{eqnarray}
where $\mathrm{Cov}_{W'}[A,B] = \mathbb{E}_{\{\bm{z}_{m}\}}[W'AB]-\mathbb{E}_{\{\bm{z}_{m}\}}[W'A]\mathbb{E}_{\{\bm{z}_{m}\}}[W'B]$ denotes the covariance under the reweighted distribution. Based on the calculated entropy, the internal energy is obtained as follows:
\begin{eqnarray}
 U&=&F+TS.
\end{eqnarray}
Figures \mbox{\ref{fig:doublewell}}(b) and (c) show the internal-energy and entropy profiles decomposed from the free-energy profile. We found that the internal energy favors a polar state because the profile has a double-well shape, and the wells become shallower as temperature increases. On the other hand, the entropic term $-TS$ favors a non-polar state because of its parabolic shape, and the profile becomes steeper as temperature increases. This indicates that the change in the free-energy profile is driven by the increase in the entropic contribution in SnTe. However, this is not surprising, because the relation $F=U-TS$ indicates that a temperature-driven phase transition reflects a competition between a structure that minimizes the internal energy ($T \rightarrow 0$) and one stabilized by entropy at high temperature. The remaining question is which atoms contribute most to the entropy.

\clearpage

To clarify the origin of entropy with atomic resolution, we visualized the entropy distribution, utilizing the fact that Eq.~(\mbox{\ref{eq:entropy}}) is a sum of per-atom entropy contributions and can be decomposed. Figure \mbox{\ref{fig:doublewell}}(d) shows the distribution of the difference in the entropic contribution in SnTe between the non-polar and polar states. We found that Sn atoms show a larger variation in entropy than Te atoms, suggesting that Sn atoms contribute more strongly to the ferroelectric-paraelectric phase transition than Te atoms. On the other hand, the entropy distribution maps (Figs.~\mbox{\ref{fig:doublewell}}(e) and \mbox{\ref{fig:doublewell}}(f)) show that Sn atoms in the non-polar state exhibit a larger atomic-vibration width elongated along the in-plane direction. Furthermore, Te atoms do not show such a remarkable change in atomic vibration, although the atomic mass of Te (127.60) is not very different from that of Sn (118.71). This behavior was also observed in previous MD simulations\mbox{\cite{PhysRevB.90.214303}}.


From these observations and previous studies, we propose the following thermodynamic picture of the phase transition:
(1) The occupied p states of Te atoms mediate the hybridization between the lone-pair s orbital of Sn and the unoccupied p orbitals of Sn\mbox{\cite{PhysRevB.67.125111,doi:10.1021/jp051822r,PhysRevLett.121.027601}}. This hybridization induces symmetry breaking in the electronic distribution and thereby stabilizes ferroelectricity in SnTe.
(2) In this ferroelectric state, the shortest interatomic distances become smaller because of the symmetry breaking. As a result, the ferroelectric state provides less space for atomic vibrations as temperature increases, and the vibrational entropy gain is therefore reduced.
(3) Consequently, the system tends to move to a state that allows larger atomic vibrations and hence a greater increase in entropy, i.e., the paraelectric state.
(4) On the other hand, considering that p-orbital overlap forms the bonding network in SnTe\mbox{\cite{doi:10.1038/s41467-024-53599-2}}, while the Sn p states are activated through hybridization, we may view the Te-centered network as the main, relatively rigid lattice framework, and the Sn atoms as a softer sublattice embedded in it. Accordingly, Sn exhibits larger atomic vibrations and hence larger vibrational entropy than Te. In summary, the ferroelectric–paraelectric phase transition can be interpreted as being driven, at least in part, by the entropic contribution of Sn atoms, which is dictated by the electronic bonding.

Finally, we outline the limitations of our model at this stage and directions for future development: (1) The above phase-transition mechanism is based on an Einstein-solid approximation; that is, we evaluate the free energy of a reference (virtual) system that approximates the free energy of the actual system. (2) We consider only vibrational entropy, although the total entropy can also include other contributions such as configurational entropy (and, for molecular systems, rotational and translational entropies). (3) The phase transition of SnTe is also influenced by the carrier density, but such effects are not considered in the present model\mbox{\cite{doi:10.1126/science.aad8609}}.

\clearpage
\section{Conclusion}
In summary, we have developed atomic-scale phase-field modeling that incorporates non-Gaussian probability densities and many-body interactions. Our modeling successfully reproduces the temperature dependence of polarization in monolayer SnTe under the NVT ensemble. Furthermore, the framework enables an atom-by-atom decomposition of the vibrational entropy. This decomposition suggests that Sn atoms contribute more to the entropy than Te atoms, which in turn drives the ferroelectric–paraelectric phase transition at high temperature. Overall, our modeling extends phase-field modeling, that is, local thermodynamic theory to the ultra-small scale, and provides an atom-resolved mechanism of the phase transition.

\begin{acknowledgement}
This work was supported by JSPS KAKENHI grant numbers 25K23433.

\end{acknowledgement}

\begin{suppinfo}
The Supporting Information is available free of charge at
\begin{itemize}
  \item PDF: Variational free energy of an anharmonic oscillator, Derivatives of the free energy, Derivation of force, Derivation of pressure, Derivation of electric polarization, Derivation of entropy, and a movie for atom dynamics.
\end{itemize}

\end{suppinfo}\textbf{}

\bibliography{manuscript}

\end{document}